\documentclass{article}
\usepackage[T1]{fontenc} % add special characters (e.g., umlaute)
\usepackage[utf8]{inputenc} % set utf-8 as default input encoding
\usepackage{ismir,amsmath,cite,url}
\usepackage{graphicx}
\usepackage{color}
\usepackage{booktabs}

\title{GAFX: A General Audio Feature eXtractor}

\threeauthors
  {Zhaoyang Bu} {Center for Safe AGI \\ Ohla AI Group\\ {\tt zhaoyangbu@arizona.edu}}
  {Hanhaodi Zhang} {Center for Safe AGI \\ Ohla AI Group\\ {\tt} hanhaodi.zhang@uqconnect.edu.au}
  {Xiaohu Zhu} {Center for Safe AGI \\ Ohla AI Group\\ {\tt xhzhu.nju@gmail.com}}

\begin{document}

\maketitle
\begin{abstract}
 Most machine learning models for audio tasks are dealing with a handcrafted feature, the spectrogram. However, it is still unknown whether the spectrogram could be replaced with deep learning based features. In this paper, we answer this question by comparing the different learnable neural networks extracting features with a successful spectrogram model and proposed a General Audio Feature eXtractor (GAFX) based on a dual U-Net (GAFX-U), ResNet (GAFX-R), and Attention (GAFX-A) modules. We design experiments to evaluate this model on the music genre classification task on the GTZAN dataset and perform a detailed ablation study of different configurations of our framework and our model GAFX-U, following the Audio Spectrogram Transformer (AST) classifier achieves competitive performance.

\end{abstract}
\section{Introduction}\label{sec:introduction}
Recently, researchers have shown an increased interest in applying successful deep learning models in computer vision (CV) and natural language processing (NLP) for the audio field. Speciﬁcally, convolutional neural networks (CNNs) architectures \cite{kong2020panns}, such as VGG \cite{simonyan2014very}, have been trained on the audio spectrograms for different tasks and one of the most used pretrained NLP model, BERT \cite{devlin2018bert}, have been transformed as AALBERT \cite{chi2021audio}, which is also pretrained on the spectrograms and fine-tuned for tasks, such as classifications. In addition, recent researches on Transformer architecture \cite{vaswani2017attention, dosovitskiy2020image, pmlr-v139-touvron21a} and Masked Auto-Encoder (MAE) \cite{he2022masked} has been studied on audio spectrograms \cite{gong2021ast, niizumi2022masked}. The spectrogram is a kind of low-level acoustic representation and the previous models usually take the mel-frequency spectrograms as inputs, which are computed from the time-domain waveform through the short-time Fourier transform (STFT) \cite{shen2018natural}. However, research on the audio field models has been mostly restricted to the hand-crafted feature inputs, which are spectrograms.

We attempt to show that learnable wave feature extractors instead of the spectrogram transform to achieve a better result on the music genre classification task. Unlike the learnable audio frontend \cite{zeghidour2021leaf} which is composed by learnable filter combination, we get inspired by how people tell the style of a song. People often distinguish the music style through its unique elements like rhythm, harmony, and melody. For example, if a song has plenty of off-beat rhythms, then its style most likely is reggae. We want our feature extractor to be able to extract the feature that contains these unique elements. We designed different feature extraction networks: GAFX-U (dual U-Net based), GAFX-R (ResNet based), and GAFX-A (Attention based) with AST as the classifier. 

This work will generate an alternative insight into the audio analysis with learnable feature extractors which produce frequency-domain-like features instead of mel-frequency spectrograms. We first introduce an End-to-End framework as a Learnable Feature Extractor and classifier in the audio waveform domain. As for the task, music genre classification, we focus on, End-to-End means the input is the original waveform and the output is the genre directly. Then, comparing the baseline model AST, the spectrogram-input model with transformer, and our designed feature extractors, proves the pretrained feature extractors have the ability to catch the features out of the hand-crafted features. Finally, the results show that the model with GAFX-U and AST classifier achieves 72.5\% accuracy and improves 11.5\% accuracy to the AST baseline.

\section{Related Work}\label{sec:related_work}
Most research on audio tasks has been classified into studies on spectrogram and waveform. In this section, previous works will be discussed in these two views.
\subsection{Methods using Spectrograms}
The researchers applied the CV succeeded models on spectrogram since spectrogram data in a certain time is the same as the 2-D image data with channel one.
\subsubsection{CNNs}
CNNs have proven very effective in audio tasks \cite{hershey2017cnn,palanisamy2020rethinking,guzhov2021esresnet,ryabinov2022comparison}. CNNs architectures have been used to classified the sound on the large-scale audio set. Researchers  \cite{hershey2017cnn,ryabinov2022comparison} first experiment several typical models, AlexNet \cite{krizhevsky2012imagenet}, VGG, Inception \cite{szegedy2016rethinking}, ResNet, MobileNetv2 \cite{sandler2018mobilenetv2} and EfficientNet \cite{tan2019efficientnet}. Then considering the same 2-D data structure of image and spectrogram, it shows that ImageNet-Pretrained \cite{deng2009imagenet} CNN models can be used as strong baseline networks for audio classification.
\subsubsection{Transformer}
The transformer is a kind of attention architecture that has been made use of for audio tasks with or without convolutional layers \cite{kong2020sound,boes2019audiovisual,miyazaki2020convolution,gong2021ast}. Transformer with CNN combining both advantages has shown the ability to efficiently capture the global and local context information of an audio feature sequence and been applied to the Sound event detection (SED) task \cite{boes2019audiovisual,miyazaki2020convolution}. On the other hand, the pure transformer, ViT \cite{dosovitskiy2020image} and its variant, DeiT\cite{pmlr-v139-touvron21a}, were proposed on the audio spectrogram, as AST models, after pretraining on the ImageNet achieved better results than CNNs with attention on the Audio Set \cite{gemmeke2017audio} tasks. 
\subsubsection{Masked Spectrogram Model}
The masked-prediction pretraining is now also applied on the spectrogram\cite{niizumi2022masked,chong2022masked}, after being applied as Masked Language Model (MLM) , BERT \cite{devlin2018bert} in NLP and Masked Auto-Encoder (MAE) \cite{he2022masked} in CV. MAE proves the Generality of the transformer and the information density of images is lower than that of texts. Same as the image, trained masked spectrogram model (MSM) \cite{niizumi2022masked} on the spectrogram with the same high mask ratio achieves good performance and at the same time masked spectrogram prediction (MaskSpec) \cite{chong2022masked} experiments different tasks on the same strategies.     

\subsection{Researches on Waveform}
One of the most differences between the waveform and spectrogram is that the hand-craft features are based on the human-hearing conception frequency filters. Therefore, the audio classification tasks labeled by people are mostly applied to the spectrogram. Correspondingly, the waveform is as the input of model for tasks, such as audio classification \cite{zeghidour2021leaf}, audio generation \cite{oord2016wavenet}, and audio separation \cite{stoller2018wave,defossez2019music,defossez2019demucs,defossez2021hybrid}. For the audio classification, a learnable audio frontend (LEAF) \cite{zeghidour2021leaf} is composed of three learnable components, filter, lowpass pooling, and compression to generate the feature which outperforms the mel-filterbank. One of the audio generation is text-to-speech (TTS), WaveNet \cite{oord2016wavenet} based on the PixelCNN \cite{van2016conditional},  Which is generative CNN, achieves state-of-the-art performance. For the audio source separation task, wave-U-Net based on the 1-D U-Net \cite{ronneberger2015u}, of which 2-D architecture used for image segmentation, generate the same result as a state-of-the-art spectrogram-based U-Net architecture.  After that, a series of Deep Extractor for Music Sources (Demucs) was proposed for the task as the name. The Demucs first add the bidirectional LSTM \cite{uhlich2017improving} models between the down-sample blocks and up-sample blocks. Then To improve the Signal-To-Distortion (SDR), the models, Hybrid Demucs separates waveform into two different auto-encoders as temporal encoder and spectral encoder.  

\subsection{Genre Classification}
Most research on genre classification has been carried out on GTZAN dataset \cite{sturm2013gtzan}. The best model based jukebox \cite{dhariwal2020jukebox} representation, which is a MLM pretrained on 1M songs \cite{castellon2021codified}.

% \begin{figure}
%  \centering
%  \includegraphics[scale=0.2]{FE.pdf}
%  \caption{The model architecture. Firstly, the input waveform through feature extractor will generate spectrogram output, giving four feature outputs. Then sent the spectrogram into AST, giving the classification results.}
%  \label{architecture}
% \end{figure}

\begin{figure*}
 \centerline{
 \includegraphics[width=1.5\columnwidth]{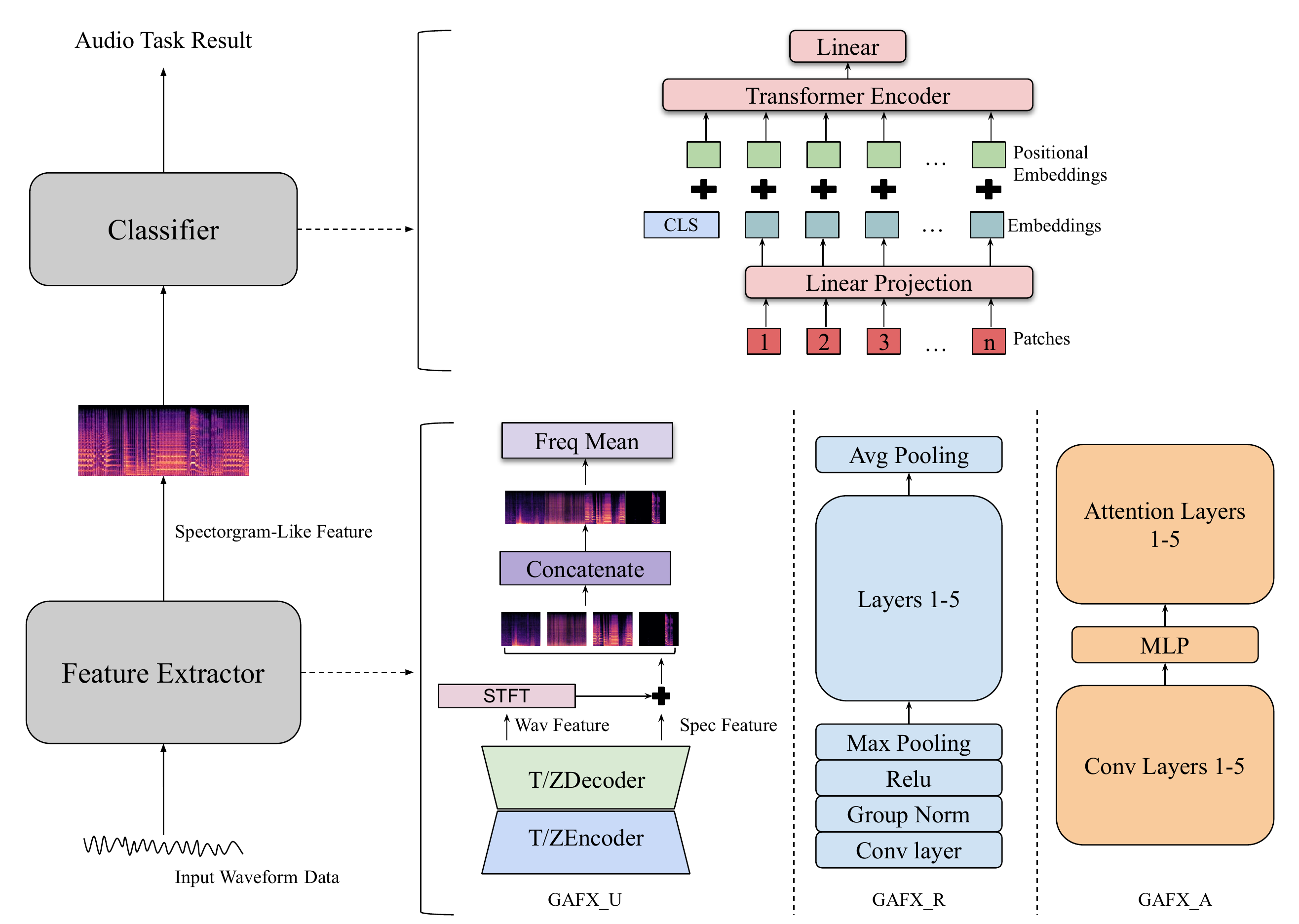}}
 \caption{The model architecture. Firstly, the input waveform through the feature extractor will generate the spectrogram-like feature. Then sent the feature into the classifier, giving the results. Details of three feature extractors and the classifier are shown on the right site of the figure.}
 \label{architecture}
\end{figure*}

\section{Method}\label{sec:method}

In this section, we will introduce the designed feature extractors and classifier for music genre classifications in detail. Feature extractors are used for extracting features directly from audio files while classifier is for distinguishing which genre the audio belongs to.

\subsection{Feature Extractor}

\begin{table}
 \begin{center}
 \begin{tabular}{ccc}
   \toprule
   Layers & TEncoder/ZEncoder & TDecoder/ZDecoder \\
   \midrule
   1 & $C_{in}$=2, $C_{out}$=48 & $C_{in}$=48, $C_{out}$=4*2*2 \\
   2 & $C_{in}$=48, $C_{out}$=96 & $C_{in}$=96, $C_{out}$=48 \\
   3 & $C_{in}$=96, $C_{out}$=192 & $C_{in}$=192, $C_{out}$=96 \\
   4 & $C_{in}$=192, $C_{out}$=384 & $C_{in}$=384, $C_{out}$=192 \\
   5 & $C_{in}$=384, $C_{out}$=768 & $C_{in}$=768, $C_{out}$=384 \\
   6 & $C_{in}$=768, $C_{out}$=1536 & $C_{in}$=1536, $C_{out}$=768 \\
   \bottomrule
 \end{tabular}
\end{center}
 \caption{The GAFX-U architecture, TEncoder and ZEncoder represents temporal encoder and spectral encoder, encoder and decoder are build symmetrically. $C_{in}$ and $C_{out}$ means numbers of input channel and output channel.}
 \label{table1}
\end{table}

\begin{table}
 \begin{center}
 \begin{tabular}{ccc}
   \toprule
   Layers &  Channels & Strides\\
   \midrule
   Conv1 & 64 & (1, 2) \\
   MaxPol & / & (1, 2) \\
   1 & 64 & (1, 1) \\
   2 & 128 & (1, 2) \\
   3 & 256 & (1, 2) \\
   4 & 512 & (1, 2) \\
   5 & 128 & (1, 2) \\
   \bottomrule
 \end{tabular}
\end{center}
 \caption{The GAFX-R architecture. For layer 1-5, the number of parameters should be twice as much as table shows since one layer contains two same blocks. }
 \label{table2}
\end{table}

\subsubsection{Feature Extractor Based on Dual U-Net}
The GAFX-U is based on the a dual U-net, redesigned Hybrid Demucs architecture (HDemucs). HDemucs is composed of two branches, a temporal branch, and a spectral branch. Each branch is like the standard Demucs, which is an encoder/decoder structure with BiLSTM between them. Since the model won the Music Demixing Challenge 2021 organized by Sony, we believe that this dual U-Net structure has abilities to catch the features behind the waveform. The HDemucs  architecture shows in table [\ref{table1}]. 

In detail, as the CV models achieve success on the spectrogram, we redesign extractors to transform the output as image-like, also spectrogram-like. Firstly, generating four waveform features from the source, STFT with 512 FFT length and hanning window are applied these features and they are added with the output of ZDecoder separately. After these steps, the four image-like features with 469 time steps and 2048 frequency bins (the shape representing as (469, 2048)). Then, these four features are concatenate along with frequency direction to one tensor with shape (469*4, 2048). Final steps, to reduce the dimension of frequency, the bins are divided into 16 groups, and each group takes the mean as representation. As a result, our GAFX-U will generate features with shape (1876, 128).

\subsubsection{Feature Extractor Based on ResNet}
The GAFX-R is based on ResNet architecture, CNN with layer connection. The structure of our GAFX-R contains 5 layers, each layer composed of 2 basic blocks in which the kernel size is the same as that of the standard ResNet\_18 block. More detail shows in table [\ref{table2}]. The shape of data changes as the following description: 

A 30 seconds 22KHz audio data has a shape (1, 661600) and firstly is reshaped to (1, 3308, 200). After the GAFX-R, the feature shape is (128, 3308, 4). The last procession for features is average pooling the last dimension and reshaping the data to (3308, 128). 

\subsubsection{Feature Extractor Based on Attention}

As the attention layers could catch the relation between the long-distance data rather than CNN, the GAFX-A is designed as attention layers followed by CNN. The output feature shape of CNN is the same as GAFX-R, (128, 3308, 4). Then these features go through the attention layers and the shape is (3308, 128).

\subsection{Classifier}
The classifier follows the steps of the original AST paper. The model is based on the image-net pretrained DeiT and with input layer weights being averaged since spectrogram with one channel rather than the RGB channels of the image. The patch size is 16*16 without overlapping because it will increase the computational complexity and the improvement in the result is not obvious. 

\section{Experiments}\label{sec:experiments}

\subsection{Dataset and Training Details}

\subsubsection{GTZAN}
The GTZAN dataset is the most-used public dataset for evaluation in machine listening research for music genre classification. GTZAN contains 1000 audio tracks divided into 10 genres with 100 tracks per genre. Each track is a 30-second single-channel wave file, with 22050 Hz sampling rate and 16-bit depth. The dataset is split into 80 percent for training and 20 percent for evaluation. Training set and evaluating set follow the same data distribution. 

For the GAFX-U we have applied data preprocessing. We first downsampled each track from 22kHz to 16kHz due to GPU memory restriction. Besides, 16kHz is enough for preserving the information we need. Then, we change each track from mono to stereo since the pretrained HDemucs model uses stereo inputs.

\subsubsection{AudioSet} 
AudioSet consists of an expanding ontology of 632 audio event classes and a collection of 2,084,320 human-labeled 10-second sound clips drawn from YouTube videos. For GAFX-R and GAFX-A we use AudioSet as a pretrain dataset. 

\subsubsection{Training details}

We train the model with the Adam optimizer, a batch size of 4. The initial learning rate is 5e-5 and the number of epochs is 80, the beginning 800 steps for warm-up. We apply the learning rate decay strategy, when the step of epoch = [12, 24, 50, 65], the learning rate will be cut in half. 

\subsection{Ablation Study}

\begin{table}
 \begin{center}
 \begin{tabular}{cc}
   \toprule
   Model & Accuracy \\
   \midrule
   AST Based on DeiT-Tiny & 61.00 \\
   AST Based on DeiT-Small & 64.50 \\
   \bottomrule
 \end{tabular}
\end{center}
 \caption{The baseline comparison of AST based on DeiT-Tiny and DeiT-Small.}
 \label{Extable1}
\end{table}

\begin{table}
 \begin{center}
 \begin{tabular}{cc}
   \toprule
   Approach & Accuracy \\
   \midrule
   w/ AudioSet Pretraining & 51.00 ± 2.00 \\
   wo/ AudioSet Pretraining & 47.10 ± 1.50 \\
   \bottomrule
 \end{tabular}
\end{center}
 \caption{The comparison of model w/ and wo/ AudioSet pretraining. In this experiment, we use GAFX-R as our feature extractor, AST with DeiT-Small architecture as our classifier, batch size is 8, and the learning rate is equal to 5e-4.}
 \label{Extable2}
\end{table}

\begin{table}
 \begin{center}
 \begin{tabular}{cc}
   \toprule
   Approach & Accuracy \\
   \midrule
   w/ Augmentation & 48.67 \\
   wo/ Augmentation & 37.00 \\
   \bottomrule
 \end{tabular}
\end{center}
 \caption{The comparison of model w/ and wo/ data augmentation. In this experiment, we use GAFX-R as our feature extractor, AST with DeiT-Small architecture as our classifier, and batch size is 4.}
 \label{Extable3}
\end{table}

\subsubsection{Baseline} 
We use pure AST architecture as our model baseline. For this part of the experiment, we first need to convert original audio inputs to mel spectrogram with 128 frequency bins and 3308 time steps. We compare the performance of the AST model based on different DeiT scales. The conclusion in the DeiT paper says that the larger DeiT model tends to achieve better classification accuracy, so the AST model may as well. we compare the accuracy between DeiT-Tiny and DeiT-Small, as the result shown in table \ref{Extable1} confirmed this conclusion, DeiT-Small outperforms 3.5 percent accuracy than DeiT-Tiny.

\subsubsection{The effect of different feature extractor} 
We compare the performance of 3 different feature extractors. The performance of GAFX-A is extremely poor, after we tried several ways to improve its performance, the loss is still NaN, so we have to abandon this feature extractor. The remaining two feature extractors are all functional. GAFX-U achieves the best result: outperforming GAFX-R with 24 percent accuracy.

\subsubsection{The effect of AudioSet pretraining} 
We compare AudioSet pretraining and using GTZAN directly. We used the balanced AudioSet for pretraining because it would generate higher accuracy models. Our strategy is to use waveform as input, and spectrogram with the shape (1024, 128) as output. We use binary cross-entropy with logits loss, Adam optimizer, the batch size is 4 and the initial learning rate is 5e-4. From the result shown in table \ref{Extable2}, we could find that AudioSet pretraining can slightly bring an improvement to the result.

\subsubsection{The effect of data augmentation} 
Data augmentation is a popular way to solve data insufficient problems, sometimes it can improve the results. We compare the GAFX-R result with the original GTZAN and augmented GTZAN. The way we augment is to split the original 30 seconds long track into three 10 seconds long tracks, so we will have triple the amount of original data. From table \ref{Extable3}, we find that data augmentation will improve accuracy.

\begin{table}
 \begin{center}
 \begin{tabular}{ccc}
   \toprule
   Feature Extractor & Accuracy \\
   \midrule
   GAFX-A & / \\
   GAFX-R & 53.00 \\
   GAFX-U & 72.50 \\
   \bottomrule
 \end{tabular}
\end{center}
\caption{Results of applying different feature extractors.}
\label{Extable4}
\end{table}

\begin{table}
 \begin{center}
 \begin{tabular}{cc}
   \toprule
   Model & Accuracy \\
   \midrule
   GAFX-U + AST(‘DeiT-Ti’) & 72.50 \\
   (Contrastive) Probing CLMR \cite{castellon2021codified} & 68.60 \\
   AST(‘DeiT-Small’) & 64.50 \\
   AST(‘DeiT-Ti’) & 61.00 \\
   \bottomrule
 \end{tabular}
\end{center}
 \caption{Results of different models for music genre classification task on GTZAN dataset.}
 \label{Extable5}
\end{table}

\section{Conclusions}

This study has shown that a designed learnable feature extractor has the ability to extract more information from waveform than spectrogram. We compared different structures of the feature extractors and strategies, such as pretraining and data augmentation with the spectrogram based AST on the music genre classification task. The research has shown that our designed extractor model achieves better results than the spectrogram baseline model. The findings of this study also suggest that the learnable feature extractors on waveform data provide the possibility of the improvements for the audio tasks. 

The generalizability of these results is subject to certain limitations. For instance, the experiments on one certain task are not enough for providing a normal the, while the feature extractors we designed are not generalized as well. Notwithstanding these limitations, the study suggests that the findings reported here shed new light on the learnable feature extractor on the waveform. 

In the future, we will replace the feature extractor and AST classifier with better result models to solve the above limitations.

\bibliography{ISMIR2022\_template}

\begin{thebibliography}{10}
\providecommand{\url}[1]{#1}
\csname url@samestyle\endcsname
\providecommand{\newblock}{\relax}
\providecommand{\bibinfo}[2]{#2}
\providecommand{\BIBentrySTDinterwordspacing}{\spaceskip=0pt\relax}
\providecommand{\BIBentryALTinterwordstretchfactor}{4}
\providecommand{\BIBentryALTinterwordspacing}{\spaceskip=\fontdimen2\font plus
\BIBentryALTinterwordstretchfactor\fontdimen3\font minus
  \fontdimen4\font\relax}
\providecommand{\BIBforeignlanguage}[2]{{%
\expandafter\ifx\csname l@#1\endcsname\relax
\typeout{** WARNING: IEEEtran.bst: No hyphenation pattern has been}%
\typeout{** loaded for the language `#1'. Using the pattern for}%
\typeout{** the default language instead.}%
\else
\language=\csname l@#1\endcsname
\fi
#2}}
\providecommand{\BIBdecl}{\relax}
\BIBdecl

\bibitem{kong2020panns}
Q.~Kong, Y.~Cao, T.~Iqbal, Y.~Wang, W.~Wang, and M.~D. Plumbley, ``Panns:
  Large-scale pretrained audio neural networks for audio pattern recognition,''
  \emph{IEEE/ACM Transactions on Audio, Speech, and Language Processing},
  vol.~28, pp. 2880--2894, 2020.

\bibitem{simonyan2014very}
K.~Simonyan and A.~Zisserman, ``Very deep convolutional networks for
  large-scale image recognition,'' \emph{arXiv preprint arXiv:1409.1556}, 2014.

\bibitem{devlin2018bert}
J.~Devlin, M.-W. Chang, K.~Lee, and K.~Toutanova, ``Bert: Pre-training of deep
  bidirectional transformers for language understanding,'' \emph{arXiv preprint
  arXiv:1810.04805}, 2018.

\bibitem{chi2021audio}
P.-H. Chi, P.-H. Chung, T.-H. Wu, C.-C. Hsieh, Y.-H. Chen, S.-W. Li, and H.-y.
  Lee, ``Audio albert: A lite bert for self-supervised learning of audio
  representation,'' in \emph{2021 IEEE Spoken Language Technology Workshop
  (SLT)}.\hskip 1em plus 0.5em minus 0.4em\relax IEEE, 2021, pp. 344--350.

\bibitem{vaswani2017attention}
A.~Vaswani, N.~Shazeer, N.~Parmar, J.~Uszkoreit, L.~Jones, A.~N. Gomez,
  {\L}.~Kaiser, and I.~Polosukhin, ``Attention is all you need,''
  \emph{Advances in neural information processing systems}, vol.~30, 2017.

\bibitem{dosovitskiy2020image}
A.~Dosovitskiy, L.~Beyer, A.~Kolesnikov, D.~Weissenborn, X.~Zhai,
  T.~Unterthiner, M.~Dehghani, M.~Minderer, G.~Heigold, S.~Gelly \emph{et~al.},
  ``An image is worth 16x16 words: Transformers for image recognition at
  scale,'' \emph{arXiv preprint arXiv:2010.11929}, 2020.

\bibitem{pmlr-v139-touvron21a}
H.~Touvron, M.~Cord, M.~Douze, F.~Massa, A.~Sablayrolles, and H.~Jegou,
  ``Training data-efficient image transformers\&amp; distillation through
  attention,'' in \emph{International Conference on Machine Learning}, vol.
  139, July 2021, pp. 10\,347--10\,357.

\bibitem{he2022masked}
K.~He, X.~Chen, S.~Xie, Y.~Li, P.~Doll{\'a}r, and R.~Girshick, ``Masked
  autoencoders are scalable vision learners,'' in \emph{Proceedings of the
  IEEE/CVF Conference on Computer Vision and Pattern Recognition}, 2022, pp.
  16\,000--16\,009.

\bibitem{gong2021ast}
Y.~Gong, Y.-A. Chung, and J.~Glass, ``Ast: Audio spectrogram transformer,''
  \emph{arXiv preprint arXiv:2104.01778}, 2021.

\bibitem{niizumi2022masked}
D.~Niizumi, D.~Takeuchi, Y.~Ohishi, N.~Harada, and K.~Kashino, ``Masked
  spectrogram modeling using masked autoencoders for learning general-purpose
  audio representation,'' \emph{arXiv preprint arXiv:2204.12260}, 2022.

\bibitem{shen2018natural}
J.~Shen, R.~Pang, R.~J. Weiss, M.~Schuster, N.~Jaitly, Z.~Yang, Z.~Chen,
  Y.~Zhang, Y.~Wang, R.~Skerrv-Ryan \emph{et~al.}, ``Natural tts synthesis by
  conditioning wavenet on mel spectrogram predictions,'' in \emph{2018 IEEE
  international conference on acoustics, speech and signal processing
  (ICASSP)}.\hskip 1em plus 0.5em minus 0.4em\relax IEEE, 2018, pp. 4779--4783.

\bibitem{zeghidour2021leaf}
N.~Zeghidour, O.~Teboul, F.~de~Chaumont~Quitry, and M.~Tagliasacchi, ``Leaf: A
  learnable frontend for audio classification,'' \emph{ICLR}, 2021.

\bibitem{hershey2017cnn}
S.~Hershey, S.~Chaudhuri, D.~P. Ellis, J.~F. Gemmeke, A.~Jansen, R.~C. Moore,
  M.~Plakal, D.~Platt, R.~A. Saurous, B.~Seybold \emph{et~al.}, ``Cnn
  architectures for large-scale audio classification,'' in \emph{2017 ieee
  international conference on acoustics, speech and signal processing
  (icassp)}.\hskip 1em plus 0.5em minus 0.4em\relax IEEE, 2017, pp. 131--135.

\bibitem{palanisamy2020rethinking}
K.~Palanisamy, D.~Singhania, and A.~Yao, ``Rethinking cnn models for audio
  classification,'' \emph{arXiv preprint arXiv:2007.11154}, 2020.

\bibitem{guzhov2021esresnet}
A.~Guzhov, F.~Raue, J.~Hees, and A.~Dengel, ``Esresnet: Environmental sound
  classification based on visual domain models,'' in \emph{2020 25th
  International Conference on Pattern Recognition (ICPR)}.\hskip 1em plus 0.5em
  minus 0.4em\relax IEEE, 2021, pp. 4933--4940.

\bibitem{ryabinov2022comparison}
A.~Ryabinov and M.~Uzdiaev, ``A comparison study of widespread cnn
  architectures for speech emotion recognition on spectrogram,'' in \emph{AIP
  Conference Proceedings}, vol. 2467, no.~1.\hskip 1em plus 0.5em minus
  0.4em\relax AIP Publishing LLC, 2022, p. 050008.

\bibitem{krizhevsky2012imagenet}
A.~Krizhevsky, I.~Sutskever, and G.~E. Hinton, ``Imagenet classification with
  deep convolutional neural networks,'' \emph{Advances in neural information
  processing systems}, vol.~25, 2012.

\bibitem{szegedy2016rethinking}
C.~Szegedy, V.~Vanhoucke, S.~Ioffe, J.~Shlens, and Z.~Wojna, ``Rethinking the
  inception architecture for computer vision,'' in \emph{Proceedings of the
  IEEE conference on computer vision and pattern recognition}, 2016, pp.
  2818--2826.

\bibitem{sandler2018mobilenetv2}
M.~Sandler, A.~Howard, M.~Zhu, A.~Zhmoginov, and L.-C. Chen, ``Mobilenetv2:
  Inverted residuals and linear bottlenecks,'' in \emph{Proceedings of the IEEE
  conference on computer vision and pattern recognition}, 2018, pp. 4510--4520.

\bibitem{tan2019efficientnet}
M.~Tan and Q.~Le, ``Efficientnet: Rethinking model scaling for convolutional
  neural networks,'' in \emph{International conference on machine
  learning}.\hskip 1em plus 0.5em minus 0.4em\relax PMLR, 2019, pp. 6105--6114.

\bibitem{deng2009imagenet}
J.~Deng, W.~Dong, R.~Socher, L.-J. Li, K.~Li, and L.~Fei-Fei, ``Imagenet: A
  large-scale hierarchical image database,'' in \emph{2009 IEEE conference on
  computer vision and pattern recognition}.\hskip 1em plus 0.5em minus
  0.4em\relax Ieee, 2009, pp. 248--255.

\bibitem{kong2020sound}
Q.~Kong, Y.~Xu, W.~Wang, and M.~D. Plumbley, ``Sound event detection of weakly
  labelled data with cnn-transformer and automatic threshold optimization,''
  \emph{IEEE/ACM Transactions on Audio, Speech, and Language Processing},
  vol.~28, pp. 2450--2460, 2020.

\bibitem{boes2019audiovisual}
W.~Boes and H.~Van~hamme, ``Audiovisual transformer architectures for
  large-scale classification and synchronization of weakly labeled audio
  events,'' in \emph{Proceedings of the 27th ACM International Conference on
  Multimedia}, 2019, pp. 1961--1969.

\bibitem{miyazaki2020convolution}
K.~Miyazaki, T.~Komatsu, T.~Hayashi, S.~Watanabe, T.~Toda, and K.~Takeda,
  ``Convolution augmented transformer for semi-supervised sound event
  detection,'' in \emph{Proc. Workshop Detection Classification Acoust. Scenes
  Events (DCASE)}, 2020, pp. 100--104.

\bibitem{gemmeke2017audio}
J.~F. Gemmeke, D.~P. Ellis, D.~Freedman, A.~Jansen, W.~Lawrence, R.~C. Moore,
  M.~Plakal, and M.~Ritter, ``Audio set: An ontology and human-labeled dataset
  for audio events,'' in \emph{2017 IEEE international conference on acoustics,
  speech and signal processing (ICASSP)}.\hskip 1em plus 0.5em minus
  0.4em\relax IEEE, 2017, pp. 776--780.

\bibitem{chong2022masked}
D.~Chong, H.~Wang, P.~Zhou, and Q.~Zeng, ``Masked spectrogram prediction for
  self-supervised audio pre-training,'' \emph{arXiv preprint arXiv:2204.12768},
  2022.

\bibitem{oord2016wavenet}
A.~v.~d. Oord, S.~Dieleman, H.~Zen, K.~Simonyan, O.~Vinyals, A.~Graves,
  N.~Kalchbrenner, A.~Senior, and K.~Kavukcuoglu, ``Wavenet: A generative model
  for raw audio,'' \emph{arXiv preprint arXiv:1609.03499}, 2016.

\bibitem{stoller2018wave}
D.~Stoller, S.~Ewert, and S.~Dixon, ``Wave-u-net: A multi-scale neural network
  for end-to-end audio source separation,'' \emph{arXiv preprint
  arXiv:1806.03185}, 2018.

\bibitem{defossez2019music}
A.~D{\'e}fossez, N.~Usunier, L.~Bottou, and F.~Bach, ``Music source separation
  in the waveform domain,'' \emph{arXiv preprint arXiv:1911.13254}, 2019.

\bibitem{defossez2019demucs}
A.~Défossez, N.~Usunier, L.~Bottou, and F.~Bach, ``Demucs: Deep extractor for
  music sources with extra unlabeled data remixed,'' \emph{arXiv preprint
  arXiv:1909.01174}, 2019.

\bibitem{defossez2021hybrid}
A.~D{\'e}fossez, ``Hybrid spectrogram and waveform source separation,''
  \emph{arXiv preprint arXiv:2111.03600}, 2021.

\bibitem{van2016conditional}
A.~Van~den Oord, N.~Kalchbrenner, L.~Espeholt, O.~Vinyals, A.~Graves
  \emph{et~al.}, ``Conditional image generation with pixelcnn decoders,''
  \emph{Advances in neural information processing systems}, vol.~29, 2016.

\bibitem{ronneberger2015u}
O.~Ronneberger, P.~Fischer, and T.~Brox, ``U-net: Convolutional networks for
  biomedical image segmentation,'' in \emph{International Conference on Medical
  image computing and computer-assisted intervention}.\hskip 1em plus 0.5em
  minus 0.4em\relax Springer, 2015, pp. 234--241.

\bibitem{uhlich2017improving}
S.~Uhlich, M.~Porcu, F.~Giron, M.~Enenkl, T.~Kemp, N.~Takahashi, and
  Y.~Mitsufuji, ``Improving music source separation based on deep neural
  networks through data augmentation and network blending,'' in \emph{2017 IEEE
  International Conference on Acoustics, Speech and Signal Processing
  (ICASSP)}.\hskip 1em plus 0.5em minus 0.4em\relax IEEE, 2017, pp. 261--265.

\bibitem{sturm2013gtzan}
B.~L. Sturm, ``The gtzan dataset: Its contents, its faults, their effects on
  evaluation, and its future use,'' \emph{arXiv preprint arXiv:1306.1461},
  2013.

\bibitem{dhariwal2020jukebox}
P.~Dhariwal, H.~Jun, C.~Payne, J.~W. Kim, A.~Radford, and I.~Sutskever,
  ``Jukebox: A generative model for music,'' \emph{arXiv preprint
  arXiv:2005.00341}, 2020.

\bibitem{castellon2021codified}
R.~Castellon, C.~Donahue, and P.~Liang, ``Codified audio language modeling
  learns useful representations for music information retrieval,'' \emph{arXiv
  preprint arXiv:2107.05677}, 2021.

\end{thebibliography}
\end{document}